\documentclass[sigconf]{acmart}

\usepackage{amsmath}
\usepackage{array} 
\usepackage[normalem]{ulem}
\usepackage{hyperref}
\usepackage{subcaption}

\AtBeginDocument{%
  }

\setcopyright{none}
\renewcommand\footnotetextcopyrightpermission[1]{}
\acmConference[KDD 2025 Workshop]{3rd Workshop on Causal Inference and Machine Learning in Practice}{August 2025}{Toronto, ON, Canada}

\begin{document}

\title{An End-to-End Pipeline for Causal ML with Continuous Treatments: An Application to Financial Decision Making}


\author{Javier Moral Hernández}
\email{javier.moral.1996@gmail.com}
\affiliation{%
  \institution{BBVA, AI Factory}
  \city{Madrid}
  \country{Spain}
}

\author{Clara Higuera-Cabañes}
\email{clara.higuera@bbva.com}
\affiliation{%
  \institution{BBVA, AI Factory}
  \city{Madrid}
  \country{Spain}
}

\author{Álvaro Ibraín}
\email{alvaroibrarod@gmail.com}
\affiliation{%
  \institution{BBVA, AI Factory}
  \city{Madrid}
  \country{Spain}
}

\renewcommand{\shortauthors}{Moral et al.}

\begin{abstract}
This paper presents an end-to-end causal machine learning (ML) pipeline designed for real-world applications with continuous treatments. The proposed framework consists of six sequential steps: dimensionality reduction, causal identification, positivity assumption violation handling, estimation, refutation and evaluation, and policy optimization. We introduce practical contributions not currently available in existing causal ML toolkits, specifically: (1) a method for detecting and quantifying positivity violations in continuous treatment settings (2) a novel, scalable two-stage dimensionality reduction framework tailored for causal inference with high-dimensional data; (3) the adaptation of sensitivity analysis and estimation methods originally designed for binary treatments to the continuous treatment space and (4) an end-to-end integration of these components into a modular, reproducible workflow. These innovations address real-world challenges in causal inference that are often not covered in theoretical frameworks but frequently encountered in industrial applications. The methodology is validated with a synthetic dataset inspired in a real-world financial debt collection use case, however its design can be applied to analogous problems across different industries. Results demonstrate that the proposed methodology offers a more computationally efficient approach and produces less biased estimates compared to standard methods for problems with continuous treatment and high-dimensional data. A fully functional GitHub repository with documented code and numbered notebooks is made available ensuring reproducibility and practical implementation. The pipeline presented is intended to contribute to closing the gap between academic approaches and practical application in industry contexts where causal ML can be highly beneficial such as the financial sector.

\end{abstract}





\maketitle

\section{Introduction}
Causal machine learning (causal ML) offers significant advantages over predictive ML, particularly in scenarios that require understanding cause-and-effect relationships, providing actionable insights, and avoiding spurious correlations. Unlike predictive ML, which identifies associations and correlations, causal ML assesses the true impact of interventions, generalize better to new scenarios, and ensure fairer, more reliable outcomes. This is especially valuable in domains where randomized control trials (RCTs) are impractical due to ethical, regulatory, operational, or economic constraints.
\cite{difficultrtc1}\cite{difficultrtc2}. 

The standard causal inference workflow includes: (1) Problem Formulation: Defining the causal question, treatment, and outcomes; (2) Identification: using causal discovery or expert-designed DAGs followed by estimand identification (e.g., backdoor, frontdoor) \cite{pearl2009causality} \cite{kunzel2019metalearners}; (3) Estimation: applying suitable methods to estimate causal effects; (4) Validation: using refutation tests, balance and weights diagnostics and sensitivity analysis to ensure robustness\cite{vanderweele2017sensitivity} and (5) Policy Optimization, where estimated causal effects and counterfactuals are used to inform decision-making and optimize interventions.

Despite its theoretical advantages, causal ML remains underutilized in industry settings\cite{markets2024causalAI} due to significant technical challenges. These include complex data structures, the intricacies of business decision-making processes, and the limitations of current causal inference frameworks when applied to industrial-scale problems.

The first critical challenge stems from the high dimensionality of modern industrial datasets that introduce a range of interconnected challenges for causal discovery and effect estimation. Key issues include:

Computational Scalability: Constraint-based algorithms like PC \cite{spirtes1991algorithm} can be efficient under sparsity, but exhibit exponential worst-case complexity \cite{kalisch2007estimating}. Score-based methods, such as Greedy Equivalence Search (GES) \cite{chickering2002optimal}, are further constrained by the super-exponential growth in Markov equivalence classes \cite{gillispie2001enumerating}, limiting their practical use as dimensionality increases.

Expert Knowledge Integration: Incorporating domain expertise in causal discovery is essential \cite{makela2022technical} but inherently subjective and labor-intensive. Experts must assess edge plausibility, resolve ambiguities, and identify spurious relationships—efforts that scale quadratically with the number of variables. Iterative refinement compounds the complexity, requiring repeated evaluations of changing graph structures.

Validation via Testable Implications: The number of required conditional independence tests grows combinatorially with dataset size, affecting both computational feasibility and statistical reliability, especially as conditioning sets increase in dimension \cite{shah2020hardness}. This also raises concerns over multiple testing and false discovery control.

Adjustment Set Selection for Estimation: Identifying valid adjustment sets becomes increasingly complex due to both the algorithmic difficulty of navigating large graphs and the exponential number of possible adjustment sets \cite{israel2023high}. This creates a tension between statistical efficiency and practical feasibility \cite{witte2020efficient}.

While much of the causal inference literature focuses on binary treatments for simplicity, many real-world decisions involve continuous treatment variables, which presents the second major challenge when using standard causal inference tools. Popular estimators like T-learners and X-learners \cite{kunzel2019metalearners} are primarily designed for binary treatments and struggle to generalize effectively to continuous settings. Similarly, sensitivity analysis tools such as E-values  \cite{vanderweele2017sensitivity} which help assess the robustness of causal estimates to unmeasured confounding, are typically not formulated to handle continuous interventions, limiting their applicability in practice. 

The third challenge in industrial applications of causal inference is the violation of the positivity assumption \cite{robins1986new}— the requirement that all individuals have a non-zero probability of receiving any treatment level. This assumption often fails in real-world settings where human decision-making limits exposure to certain treatments, and randomized control trials are not feasible. Early work by Petersen et al.~\cite{petersen2010diagnosing} provided a comprehensive framework for diagnoses and remedies for positivity violations but it is mainly focused on binary treatments. More recent approaches by Guo et al. \cite{guo2021violations} propose solutions to continuous treatments using techniques such as local density estimation, trimming, and weighting to address regions of low treatment overlap.

Lastly, while existing Python frameworks provide valuable methods for addressing individual challenges in causal inference (e.g., DoWhy\cite{dowhypaper}, causalml\cite{causalml}, EconML\cite{econml}), there remains a significant gap in comprehensive methodological frameworks that address the aforementioned challenges encountered in industrial applications.  For instance, DoWhy is not prepared for continuous outcome or high dimensional data and Microsoft EconML library offers tools for heterogeneous treatment effect estimation but primarily focuses on estimation rather than the complete pipeline from data preparation to policy optimization. There exist limited work on pipelines that simultaneously handle high-dimensional data, continuous treatments, positivity violations and few data samples while providing practical guidance for implementation in real settings.

The financial sector stands to gain significantly from causal methods, which reveal cause-and-effect relationships beyond mere correlations. In \cite{gupta2023causal} authors provide a comprehensive overview of the application of causal inference methods in the banking, finance, and insurance sectors. These methods have been applied in areas like investment management \cite{athey2019ml}, fraud detection \cite{zhang2024causalfd}, and fair credit decisions \cite{kilbertus2017discrimination}. However, despite growing interest, the field lacks standardized frameworks and example applications suited to finance. The complexity of financial data, such as heterogeneity in client profiles and sparsity in intervention histories—further, and the wide range of available techniques pose challenges to effective implementation in real settings.

This paper introduces a novel end-to-end causal machine learning framework designed to overcome the afore mentioned challenges. Validated on a financial debt collections use case and a synthetic dataset. The work aims to promote the use of causal methods in financial applications and other fields in which operational decision-making is critical. The paper is structured as follows. First it outlines the study’s scope and main contribution, followed by description of the experimental design, presentation of results, and concluding with a discussion of findings, limitations, and future research directions.

The code for reproducing all experiments is publicly available at
\href{https://github.com/javiermoralh/causal-pipeline}{\texttt{github.com/javiermoralh/causal-pipeline}}.
\section{Problem statement / Case study}
\label{sec:problem_statement}
In banking, the debt collections department manages clients in loan default using strategies like refinancing, debt sales, or write-downs—partial debt forgiveness in exchange for immediate repayment. While some clients repay successfully after a write-down, others do not, making optimal decision-making crucial.

Traditionally, such decisions rely on expert judgment to balance repayment likelihood and loss minimization. While automation could enhance this process, prediction alone is insufficient— it demands a causal understanding of how different write-down levels affect repayment outcomes.

Though randomized controlled trials (RCTs) would offer ideal estimates of causal effects, they are impractical in this context due to ethical concerns and regulatory barriers. Practitioners must instead rely on observational data, which is subject to confounding bias: clients in worse financial health typically receive larger reductions, making it difficult to isolate causal effects.

Given the infeasibility of RCTs, the presence of confounding, and the need for personalized, counterfactual insights, causal inference methods become essential for optimizing debt collection strategies effectively and fairly.

\section{Methods}

In order to address the aforementioned challenges we propose a series of steps in a form of a pipeline that practitioners can use in order to solve domain agnostic problems as long as any of the above issues are present. The pipeline is comprised of six steps that can be run sequentially. These steps are depicted in Figure \ref{fig:pipelinediagram} in Appendix \ref{sec:app-figures}.

\begin{enumerate}
\item \textbf{Dimensionality Reduction}: A structured approach to identify potential confounders and outcome-only causes from high-dimensional data.
\item \textbf{Identification}: A hybrid method combining algorithmic discovery with domain knowledge to get the final adjustment set.
\item \textbf{Positivity Assumption Violation Handling}: A framework for detecting and quantifying regions of limited overlap in treatment assignment, alongside remediation strategies to address identified violations.
\item \textbf{Estimation}: Methods for continuous treatment effects estimation.
\item \textbf{Refutation and Evaluation}: Comprehensive evaluation techniques for estimate selection.
\item \textbf{Policy Optimization}: Methodology for deriving optimal decision policies.
\end{enumerate}

This work assumes that the causal estimand of interest is identifiable through the backdoor criterion \cite{pearl2009causality}. Consequently, every stage of the pipeline – most notably the dimensionality-reduction and identification procedures – are tailored to selecting and balancing only those covariates that close backdoor paths (i.e., confounders). This focus implies that practitioners need control merely for these backdoor variables, simplifying the analysis while underscoring the importance of correctly detecting them to avoid residual bias.

In the following subsections each one of this blocks is covered in more detail.

\subsection{Dimensionality Reduction}
\label{dimensionality_reduction}
Applying causal inference to high-dimensional data requires a reduction in dimensionality that maintains causal integrity while addressing computational limits and potential positivity violations. Let $\mathcal{D} = {(X_i, T_i, Y_i)}_{i=1}^N$ denote the dataset, where $X \in \mathbb{R}^d$ are covariates, $T \in \mathbb{R}$ is the treatment, and $Y$ is the outcome. We propose a two-stage selection framework to construct a reduced adjustment set $\mathbf{Z} = \mathbf{Z}_T \cup \mathbf{Z}_Y$, where $\mathbf{Z}_T$ captures treatment predictors and $\mathbf{Z}_Y$ includes outcome-relevant covariates. This reduction aims to provide a manageable set of candidate adjustment variables for the identification phase, which further refines them using causal criteria and domain expertise to find the backdoor variables. 

In Stage 1, $\mathbf{Z}_T$ is identified using predictive machine learning-based feature selection to capture variables predictive of treatment assignment. Stage 2 uses dual partial-correlation analysis to identify covariates that influence the treatment–outcome relationship and covariates that affect only the outcome. The first type of covariates are retained as candidate \emph{confounders}—they must be controlled for to obtain an unbiased effect under the backdoor criterion—whereas the second type, while not mandatory, are desirable because their inclusion can lower the variance of the causal estimate \citep{facure2023neutral,brookhart2006variable}. The first subset in $\mathbf{Z}_Y$ is identified by computing the partial correlation between $T$ and $Y$, conditional on each covariate $X_j$:

\begin{equation}
\rho(T, Y|X_j) = \text{Corr}(\text{Res}(T \sim X_j), \text{Res}(Y \sim X_j))
\end{equation}

where $\text{Res}(\cdot)$ denotes regression residuals. Covariates inducing substantial deviations from the baseline correlation $\rho(T, Y)$ are retained as potential confounders. For outcome-only predictors, variables strongly correlated with $Y$ independent of $T$—determined via $\rho(X_j, Y|T)$—are included in $\mathbf{Z}_Y$.

The framework employs distinct approaches for treatment-related $\mathbf{Z}_T$  and outcome-related $\mathbf{Z}_Y$ selection, reflecting fundamental differences in how confounding and positivity violations affect these relationships. The framework distinguishes between treatment- and outcome-related features due to differing vulnerability to confounding and positivity violations. Standard predictive machine-learning feature selection can reliably capture treatment-related covariates because, in most observational settings, treatment assignment follows systematic decision rules that produce stable covariate–treatment links even where overlap is sparse \citep{mccaffrey2004propensity,schneeweiss2009hdps}. However, for outcome-related features, standard feature selection methods become unreliable due to the interplay between positivity violations and confounding. In regions where certain covariate-treatment combinations are unobserved, traditional feature selection techniques may identify spurious correlations that exist in the observed regions but do not represent genuine causal relationships. The dual partial correlation analysis circumvents these challenges by explicitly conditioning on covariates when examining confounders selection and conditioning on treatment when examining outcome-only predictor relationships.

This approach enables scalable, causally valid inference by reducing dimensionality while preserving essential confounding structures.

 \subsection{Identification}
 \label{sec:identification}
Following dimensionality reduction, the identification phase integrates algorithmic causal discovery with domain expertise validation. As Mäkelä et al. (2022) \cite{makela2022technical} highlight, causal discovery algorithms generate hypotheses rather than definitive conclusions, necessitating expert refinement. This phase employs an ensemble of methods—PC \cite{spirtes1991algorithm}, FCI \cite{spirtes1995causal}, and GES \cite{chickering2002optimal}—leveraging their complementary strengths to increase confidence in consistently identified relationships.

Algorithmic outputs serve as initial structural hypotheses, iteratively refined through domain expertise:

\begin{itemize} 
    \item Temporal Constraints: Ensuring relationships align with known variable orderings.
    \item Edge Validation: Assessing plausibility and temporal consistency.
    \item Causal Direction: Resolving algorithmic uncertainty with expert knowledge.
    \item Missing Relationships: Identifying overlooked causal links.
    \item Spurious Correlations: Removing statistically significant but non-causal associations.
\end{itemize}

Identifying and removing spurious correlations remains a key challenge, as statistical associations may lack theoretical justification. Experts assess these cases, refining the graph through additional statistical tests and confounder control. The complexity of this process scales quadratically with graph size, reinforcing the importance of prior dimensionality reduction.

Final causal graphs undergo rigorous validation via testable implications (conditional independence tests) until statistical and expert criteria align. The refined graph determines the adjustment set for causal effect estimation using identification algorithms such as the backdoor criterion \cite{pearl2009causality}, with outcome-related variables included to enhance precision \cite{facure2023neutral}.

\subsection{Positivity Assumption Violation Quantification}
\label{sec:positivivity}

We propose a three-step, model-agnostic procedure to detect regions of the covariate space where lack of overlap violates the positivity assumption for a continuous treatment based on Hirano et al. (2004) framework \cite{hirano2004propensity}.

\textbf{1. Treatment prediction.}  
Fit a flexible regression $f\!:\!Z\!\to\!\mathbb{R}$ that predicts $T$ and form residuals $\varepsilon = T-f(Z)$.

\textbf{2. Residual-density estimation.}  
Estimate the conditional density of the residuals $g(\varepsilon\mid Z)$ with any consistent method, e.g.

\begin{enumerate}
  \item kernel density estimators with data-driven bandwidth or plug-in rules  \cite{silverman1986density, scott1992multivariate}
  \item a homoscedastic Gaussian with analytic variance  
  \item conditional normalizing flows such as the GOALDeR continuous-GPS estimator, which learn $g(\varepsilon\mid Z)$ via invertible neural networks and naturally handle heteroscedasticity \cite{durkan2019neural,gao2025goalder}.  
\end{enumerate}  

The preferred model can be selected by held-out (pseudo)-log-likelihood or PSIS-LOO.

\textbf{3. Overlap quantification.}  
For any interval $[t_1,t_2]$ compute the generalised propensity score  
\[
P\!\bigl(T\!\in[t_1,t_2]\mid Z=z\bigr)=\int_{t_1}^{t_2} g\!\bigl(t-f(z)\mid z\bigr)\,dt.
\]  
A practical violation is flagged whenever this probability falls below a small tolerance $\epsilon$. The scalar $P(T\!\in[t_1,t_2]\mid Z)$ offers a continuous measure of violation severity, highlights covariate regions with poor support, and guides trimming or re-weighting—without tying practitioners to a single residual-density specification.

Common remediation strategies following \citet{petersen2010diagnosing} include
(i) \textbf{covariate restriction} to drop variables that force non-overlap,
(ii) \textbf{trimming} units whose generalised propensity scores lie in the tails,
(iii) \textbf{weight stabilisation / truncation} to temper extreme inverse probabilities,
(iv) \textbf{design modification or data augmentation} to avoid unlikely treatment levels, and
(v) \textbf{model-based extrapolation} when strong substantive knowledge supports it.
Each option trades bias for variance: trimming lowers power, reweighting inflates uncertainty, and extrapolation leans on unverifiable assumptions.

Before deciding on a remedy, analysts should check overlap quality by using some diagnostic tools as 
(a) checking Standardised Mean Differences of each covariate across treatment strata or after weighting \cite{rosenbaum1985constructing};
(b) plotting the GPS density to flag regions with large density-ratios ;
(c) inspecting the distribution of stabilised inverse-probability weights for extreme values; and
(d) verifying a common support region \cite{rosenbaum1983centralrole} via overlayed treatment densities.

Given these challenges, the objective shifts from eliminating violations to minimizing bias while maintaining practical utility. This structured positivity violation framework bridges theoretical causal inference with real-world application, enabling practitioners to assess the validity and generalizability of causal estimates.  

\subsection{Estimation}
\label{sec:estimation}

With the adjustment set $Z$ in place, the goal is to recover the conditional average \emph{dose–response curves}—i.e.\ $\mathbb{E}[Y(t)\mid Z_i]$ for every feasible treatment level $t$.  
Unlike binary interventions, continuous treatments require modelling this entire surface, typically by evaluating potential outcomes at several grid points \cite{rubin1974estimating}.

We benchmark three representative estimators (others can be swapped in):

\begin{enumerate}
    \item \textbf{Regression adjustment with interactions}.  
    A linear (or logistic) model of $Y$ on $(T,Z)$ yields an interpretable baseline; $T$’s coefficient and its interactions capture average and heterogeneous effects \cite{pearl2009causality}.
    
    \item \textbf{S-learner} \cite{kunzel2019metalearners}.  
    A single flexible learner fits $Y=f(T,Z)$. This method can apply more advanced regularization techniques in order to capture complex relationships while preventing overfitting, enabling the identification of non-linear interactions and heterogeneous effects.
    
    \item \textbf{Augmented IPTW for continuous $T$} \cite{kurz2021augmented}. This approach extends the classical AIPTW methodology to accommodate continuous treatment scenarios. Generalized Propensity Scores produce inverse-probability weights that, combined with an outcome model, form a \emph{doubly-robust} estimator—consistent if either component is correct.
\end{enumerate}

Together, these methods span a spectrum from simple, transparent baselines to more expressive and robust machine-learning approaches.

\subsection{Refutation and Evaluation}
\label{ref_eval_section}

This step is aimed at detecting potentially invalid estimates in order to avoid, as much as possible, spurious relationships that can remain after the aforementioned steps. The choice of refutation methods, such as the ones presented below, is open to the practitioner. However, in the scope of this work we only study the effect of the following two: (1) \textbf{Placebo Treatment Replacement test} \cite{dowhypaper}: With this test, actual treatments are substituted with random variables following identical distributions. Here, valid estimates should demonstrate no effect on outcomes, manifested as flat effect curves. (2) \textbf{Random Common Cause test} \cite{dowhypaper}: Evaluates estimate stability by introducing synthetic random confounders unrelated to both treatment and outcome variables. When successful, this test must yield consistent causal curves before and after the introduction of random confounders to the adjustment set.

Residual bias attributable to unmeasured confounders may persist even after the refutation procedures described above.  
To quantify the extent to which such bias could attenuate or overturn the estimated effects, we implement a sensitivity analysis based on the \emph{E-value} \cite{vanderweele2017sensitivity} framework. The \emph{E-value} quantifies the minimum strength of unmeasured confounding necessary to nullify the estimated effects. Higher E-values indicate greater estimate robustness. For continuous treatments, we extend the E-value methodology by computing a dose-response sensitivity curve $E(t)$, where for each treatment level $t$, we calculate the E-value required to nullify the estimated effect for a unit change in exposure centered at $t$.

Finally, estimator performance in ranking individuals by anticipated uplift is assessed with the Qini curve \cite{radcliffe2007using}, which plots cumulative gain against the cumulative proportion of treated units ordered by the model.  The area under this curve (AUC\textsubscript{Qini}) condenses the curve into a single, sample-size-invariant metric, thereby enabling rigorous comparison across competing estimators.

Collectively, these three assessments—falsification tests, quantitative sensitivity metrics, and uplift-oriented performance curves constitute a concise yet rigorous validation suite for continuous-treatment causal inference.

\section{Experimental setup}

To evaluate the validity of the proposed method, we assess its performance within a financial context where the objective is to determine the optimal amount of debt write-downs that minimizes the total expected loss.  Given the sensitive nature of financial data, a real publicly available version cannot be provided for reproducibility. Instead, a synthetic dataset has been generated that captures the essential characteristics and statistical properties of the original data, while not being an exact replication. 

Specifically, the data generation process incorporates: (i) a continuous treatment variable representing debt loss percentages in the $[0, 100]$ range, (ii) non-linear and heterogeneous treatment effects through carefully crafted probability functions with interaction terms, (iii) systematic confounding bias through structured covariate relationships, (iv) high dimensionality with $410$ features including both relevant causal variables and noisy ones, and (v) deliberate positivity assumption violations through a logistic-based assignment mechanism that creates regions of limited overlap in treatment assignment. The performance of the proposed method is then analyzed for the synthetic setting. More details on the generation process and its characteristics are explained in Appendix \ref{data_generation}.

Evaluation targets three aspects: (i) recovery of the true causal covariates, (ii) accuracy of conditional average dose-response curves which are shown in Figure \ref{fig:figure3A}, and (iii) computational efficiency.

Causal-variable retrieval is scored with precision and recall against the known adjustment set. Dose-response estimation bias is quantified by the root-mean-squared error (RMSE) between the true and estimated curves, averaged across 100 treatment levels on all validation cases. Specifically, for each individual $i$ in the validation set, we compute both the ``true'' conditional average dose-response curve $f_i(t)$–coming from our generation process as $\mathbb{E}[Y(t)\mid Z_i]$–and the estimated curve $\hat{f}_i(t)$ at discrete treatment levels $t \in \{1,2,\ldots,100\}$, corresponding to the percentage of debt loss. The RMSE is then computed at each treatment level across all individuals to obtain the individual bias $\mathcal{B}(t)$:

\begin{equation}
    \mathcal{B}(t) = \sqrt{\frac{1}{n}\sum_{i=1}^n(f_i(t) - \hat{f}_i(t))^2}
\end{equation}

The final metric, $\overline{\mathcal{B}}$, is obtained by averaging all individual biases over the entire evaluation set. Estimators that fail placebo or random-common-cause tests are discarded. Subsequently, the remaining methodologies are comparatively assessed through E-value sensitivity analysis and Qini-AUC in order to determine optimal performance characteristics.

Finally, we conduct an ablation study by disabling the dimensionality reduction and causal discovery components in the pipeline. This analysis isolates the marginal contribution of each step to the overall performance, providing insights into their impact on causal identification, estimation accuracy, and computational efficiency.

A comprehensive specification of the method's configuration, including correlation thresholds for dimensionality reduction, algorithmic choices for treatment-predictive feature derivation, and parameters for positivity violation detection, is provided in Appendix \ref{pipeline_configuration}. This documentation ensures methodological transparency and facilitates reproducibility.

\section{Results}

This section presents the results obtained at each stage of the proposed framework. We begin by examining improvements in causal variable selection, including a runtime analysis to assess computational efficiency. We then evaluate the enhancements in causal effect estimation and analyze the outputs that support informed decision-making in the proposed case study.

\subsection{Causal identification improvement}

The evaluation of the dimensionality reduction (3.1) and identification (3.2) phases reveals the effectiveness of the proposed approach in correctly identifying control variables for causal effect estimation within a high-dimensional feature space. Table \ref{tab:identification_results} presents a comparative analysis between the proposed methodology and the baseline approach

\begin{table}
\centering
\caption{Comparison of Variable Identification Performance}
\label{identification_results}
\label{tab:identification_results}
\begin{tabular}{>{\arraybackslash}m{2.1cm}  >{\centering\arraybackslash}m{1.1cm}  >{\centering\arraybackslash}m{1.0cm}  >{\centering\arraybackslash}m{1.3cm}  >{\centering\arraybackslash}m{1.0cm} }

\hline
\textbf{Method} & \textbf{Precision} & \textbf{Recall} & \textbf{Covariates} & \textbf{Runtime (min)}\\
\hline
Baseline & 0.05 & 1 & 169 & >600 \\
Dim. Reduction & 0.53 & 1 & 15 & 3 \\
Dim. Reduction \& Identification & 0.80 & 1 & 10 & 4 \\
\hline
\end{tabular}
\label{tab:selection_metrics}
\end{table}

The baseline approach achieved a precision of $0.05$ with perfect recall ($1.0$), selecting 169 covariates—typical of traditional causal discovery methods applied to high-dimensional data. Among the three algorithms tested (PC, FCI, GES), only PC completed execution within 600 minutes, making it the basis for baseline results.

Our two-stage methodology significantly improved performance. Dimensionality reduction increased precision to $0.53$ while maintaining recall at $1.0$, reducing the covariate set to 15. The final identification phase further improved precision to $0.80$, yielding a set of 10 covariates, including all 8 true causal controls and 2 treatment-related variables. While such variables can inflate estimator variance without reducing bias \cite{facure2023neutral}, their inclusion remains a known challenge in causal inference.

In terms of computational efficiency, the baseline required over 600 minutes for causal discovery, whereas the proposed approach reduced total runtime to just 4 minutes (3 minutes for dimensionality reduction and 1 minute for identification). This improvement is critical for industrial applications, enabling rapid iteration and expert feedback integration.

\subsection{Estimates Performance}
The evaluation of treatment effect estimation bias reveals significant improvements through the proposed methodology. Table \ref{tab:bias_comparison} presents the comparative analysis of mean averaged bias across the three evaluated approaches. The baseline methodology utilizing the full feature set exhibits the highest estimation bias ($\overline{\mathcal{B}} = 0.292$, 95\% CI $[0.279, 0.303]$), demonstrating the detrimental impact of high-dimensional noise and spurious correlations. Restricting the adjustment set to the identified causal controls while omitting synthetic data augmentation reduces bias by 19\% ($\overline{\mathcal{B}} = 0.236$, 95\% CI $[0.212, 0.257]$), though residual bias persists due to unaddressed positivity violations. The complete proposed methodology achieves superior performance with $\overline{\mathcal{B}} = 0.117$ (95\% CI [0.105, 0.131]), representing a 60\% reduction in mean bias compared to the baseline.

Among the three estimation methodologies described in section \ref{sec:estimation}, the S-learner emerged as the optimal approach. Initial refutation testing conducted with $100$ bootstrapped samples revealed significant limitations in the AIPTW estimator, which failed to pass the placebo treatment replacement test (Figure \ref{fig:figure5A} in Appendix \ref{sec:app-figures}) at extreme debt loss percentages where the dose-response curve should theoretically approach zero. 

Consequently, subsequent sensitivity analyses through E-values (Figure \ref{fig:figure10A} in Appendix \ref{sec:app-figures}) and cumulative gain curves (Figure \ref{fig:figure11A} in Appendix \ref{sec:app-figures}) focused exclusively on the S-learner and Linear Regression approaches. While both methodologies demonstrated comparable robustness in E-value analysis, the S-learner exhibited markedly superior performance in the AUC curve evaluation, ultimately justifying its selection as the primary estimation methodology for the comprehensive ablation study.

\begin{table}[ht]
\centering
\caption{Comparison of Estimation Bias Across Methodologies}
\label{tab:bias_comparison}
\begin{tabular}{lcc}
\hline
Methodology & Mean Bias & 95\% Confidence Interval \\ \hline
Baseline & 0.292 & [0.279, 0.303] \\
Adjustment Set Only & 0.236 & [0.212, 0.257] \\
Proposed Methodology & 0.117 & [0.105, 0.131] \\ \hline
\label{tab:biascomparison}
\end{tabular}
\end{table}

The treatment-level bias analysis, visualized in Figure~\ref{fig:figure1A} in Appendix \ref{sec:app-figures}, demonstrates consistent superiority of the proposed methodology across the entire treatment domain. While all methods exhibit increased bias at extreme treatment values ($0-20\%$ and $80-100\%$ debt loss), the proposed approach shows particular robustness in these regions with maximum RMSE of 0.20 compared to 0.5 for the baseline. This enhanced performance is attributed to the monotonic synthetic data augmentation strategy that effectively mitigates positivity violations in low-density treatment regions. The methodology maintains stable estimation quality across mid-range treatments ($20-80\%$) with RMSE consistently below $0.20$, significantly outperforming both baseline approaches.

This enhanced precision results from the synergistic combination of dimensionality reduction's variance minimization and synthetic augmentation's support expansion. The ablation study reveals that $32\%$ of total bias reduction originates from proper adjustment set identification, while $68\%$ derives from addressing positivity violations through domain-informed synthetic data generation.

These results suggest that while complete elimination of estimation bias remains challenging in practical applications with extreme positivity violations, the proposed framework achieves substantial improvements over conventional approaches. 

\subsection{Case Study Solution}

The application of the proposed pipeline to the debt collections case study yields two primary analytical outputs that enable informed decision-making regarding optimal write-down levels. First, the estimated causal DAG (Figure~\ref{fig:figure7A} in Appendix \ref{sec:app-figures}) reveals the underlying structural relationships between financial indicators, demonstrating that the debt treatment assignment (write-down percentage) is influenced by multiple customer characteristics. This structural understanding validates the necessity of controlling for these confounding variables to obtain unbiased treatment effect estimates.

Second, the conditional average dose-response curves (Figure~\ref{fig:figure8A} in Appendix \ref{sec:app-figures}) illustrate the heterogeneous causal effects of different write-down percentages on repayment probability across the customer population. These curves represent counterfactual predictions for each customer under varying treatment intensities, enabling precise personalization of write-down offers. The substantial variation in curve shapes and slopes indicates marked heterogeneity in treatment effects, suggesting that uniform write-down policies would be suboptimal.

These estimated potential outcomes provide the foundation for subsequent policy optimizations, where institution-specific cost functions can be applied to determine optimal write-down levels that balance recovery probability against financial loss.

\section{Discussion and Future Work}

The proposed method addresses key challenges in applying causal machine learning to industrial contexts—challenges that are often overlooked in existing frameworks—including high-dimensional data, positivity assumption violations, and continuous treatments. It introduces an end-to-end, adaptable methodology, validated through a financial debt collection use case, and is designed for broader applicability across industries facing similar constraints

Despite its strengths, the pipeline has limitations that warrant further exploration: Dimensionality reduction currently relies on linear partial correlation, which may miss non-linear relationships and interaction effects; future improvements could incorporate mutual information or kernel-based methods. Positivity violation detection, based on GPS and kernel density estimation, could be enhanced to account for unobserved confounding. The data augmentation strategy assumes a monotonic treatment-response, which may not hold in all cases and could require adaptation for non-monotonic scenarios.

These limitations point to promising directions for future research, aimed at refining the framework and extending its utility in complex, real-world applications.

\begin{acks}

Authors would like to thank Dr. Ramin Ghelichi for his generous guidance, thoughtful insights, and unwavering support throughout this work. Authors also extend their gratitude to BBVA AI Factory for its continued support of research initiatives and Ricardo García for his valuable guidance from BBVA Advanced Analytics Discipline and Risk Analytics Department. Special thanks to GRM and Risk Collections teams for their contributions and expertise, which helped keep the methodology grounded in real-world needs.

\end{acks}

\bibliographystyle{ACM-Reference-Format}
\bibliography{sample-base}

\clearpage

\section*{Appendix}
\appendix

\section{Synthetic Data-Generation}
\label{data_generation}
To validate the proposed methodological framework, we conducted experimentation using synthetic data. This approach enables rigorous evaluation of the pipeline's effectiveness under controlled conditions where the true causal relationships and treatment effects are known. The synthetic dataset was carefully designed to emulate the characteristics and challenges encountered in real-world financial applications while incorporating specific structural properties that test the pipeline's capabilities.

The dataset comprises covariates sampled from probability distributions approximating real financial indicators, including variables related to credit history and debt profiles. These covariates are categorized into three distinct groups: five confounders influencing both treatment assignment and outcome, three outcome-only variables directly affecting repayment probability but not treatment decisions, and two treatment-only variables predictive of debt loss levels but unrelated to the outcome. This structure ensures a realistic confounding scenario where treatment assignment is systematically biased by variables that also drive repayment outcomes. 

The continuous treatment variable, representing the percentage of debt loss, is generated through a non-linear function of the confounders and treatment-only variables \eqref{eq:treatment}. The treatment assignment mechanism is deliberately constructed to systematically violate the positivity assumption. This is implemented through a logistic-based assignment function with non-linear interactions between covariates, followed by scaling to the [0,100] range. While modest random noise is incorporated via truncated normal distribution to maintain realism, the underlying structure ensures that specific regions of the treatment space become practically inaccessible for certain covariate profiles. This design choice authentically mirrors real-world scenarios where financial advisors assign writedown levels based on rigid policy rules tied to customers’ financial health, creating covariate strata with no overlap across treatment ranges. 

The outcome generation process was engineered to exhibit heterogeneous treatment effects as visually shown in Figure \ref{fig:figure3A} and \ref{fig:figure15A} in Appendix \ref{sec:app-figures}, where the impact of interventions varies substantially across different covariate profiles. This heterogeneity is achieved through a carefully crafted probability function that incorporates both direct effects and interaction terms between treatments and covariates \eqref{eq:outcome}. Concretely, the inclusion of quadratic terms and multiplicative interactions between covariates guarantees heterogeneous treatment effects across individuals, with each customer exhibiting a unique dose-response curve. When combined with the treatment assignment mechanism, this generates a clear Simpson's paradox as demonstrated in Figure \ref{fig:figure2A} in Appendix \ref{sec:app-figures}, where the real average treatment-outcome relationship differs markedly from the observed outcome average at each treatment level. 

\begin{figure}[H]
    \centering
    \includegraphics[width=0.4\textwidth]{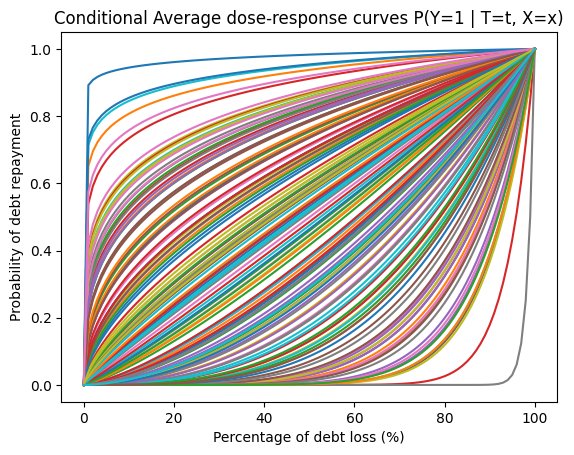}
    \caption{Heterogeneous treatment effects in synthetic generated data - P(Y=1|T=t, X=x)}
    \label{fig:figure3A}
\end{figure}

\begin{figure}[H]
    \centering
    \includegraphics[width=0.4\textwidth]{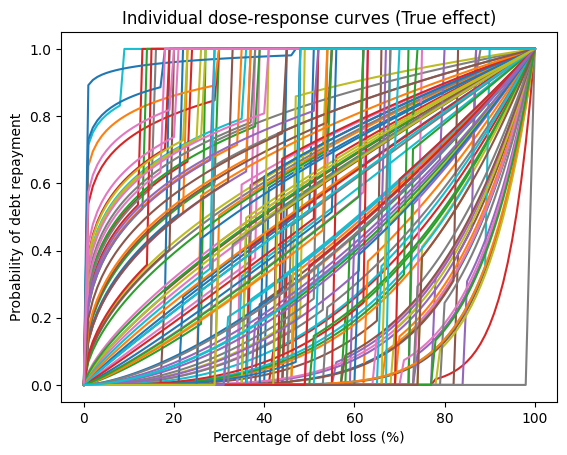}
    \caption{Heterogeneous treatment effects in synthetic generated data - Observed individual outcomes}
    \label{fig:figure15A}
\end{figure}

\begin{figure}[H]
    \centering
    \includegraphics[width=0.4\textwidth]{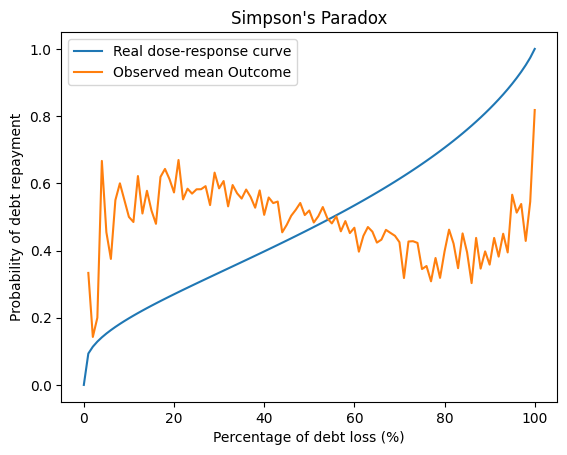}
    \caption{Simpson's paradox in synthetic generated data.}
    \label{fig:figure2A}
\end{figure}

To further approximate real-world data complexity, the feature space was augmented with two additional categories of variables. First, 100 redundant features were generated by introducing controlled correlations with the existing causal variables, maintaining separate correlation structures for confounders, treatment-only, and outcome-only predictors. Second, 300 pure noise features were synthesized by sampling from various probability distributions commonly observed in financial data, ensuring these variables maintain no systematic relationship with either the treatment or outcome. This enriched feature space creates a challenging environment for causal discovery and effect estimation, testing the pipeline's capability to identify relevant variables and discard spurious relationships.

The synthetic data generation process thus encapsulates the core challenges outlined in Section \ref{sec:problem_statement}: high dimensionality, positivity violations, continuous treatments, and heterogeneous effects. By design, it provides a controlled environment to evaluate the pipeline’s ability to recover ground truth causal relationships, isolate confounding bias, and estimate personalized treatment effects—all while mirroring the statistical complexities of real financial data.

\section{Pipeline configuration}
\label{pipeline_configuration}
\subsection{Dimensionality Reduction}
The dimensionality reduction phase was implemented through the two-stage feature selection process explained in section \ref{dimensionality_reduction}. The first stage utilized a hybrid approach combining Fast Correlation-Based Filter (FCBF) with Sequential Forward Selection (SFS) to identify variables predictive of treatment assignment. The FCBF threshold was set to $\delta = 0.0001$, ensuring the elimination of redundant features while preserving variables with substantial pairwise correlations to the treatment. The subsequent SFS phase employed a CatBoost regressor (hyperparameters are available in the code repository). A $K=3$-fold cross-validation framework with root mean squared error (RMSE) minimization guided feature inclusion,  terminating when adding a feature no longer reduced the cross-validated RMSE. The second stage refined confounder detection and outcome prediction through dual partial correlation analyses. For confounder identification, variables altering the treatment-outcome relationship were retained using a threshold of $\rho_{\text{min}} = 0.01$ on the partial correlation difference $|\rho(T,Y|X_j) - \rho(T,Y)|$. Concurrently, outcome predictors were selected under a stricter threshold of $\rho_{\text{min}} = 0.1$ for $\rho(X_j,Y|T)$, ensuring robust associations independent of treatment effects. Both analyses applied a correlation threshold of $0.5$ to eliminate multicollinear features, aligning with the synthetic data’s redundancy structure. This dual-threshold approach balanced sensitivity to weak confounders with computational efficiency, critical for scalability in industrial applications. 

\subsection{Identification}
While the methodology presented in Section \ref{sec:identification} emphasizes the critical role of domain expertise in causal discovery and graph refinement for real-world applications, the synthetic nature of the experimental dataset necessitates a modified approach to maintain experimental validity. In this controlled setting, domain knowledge application is deliberately constrained to avoid inadvertently leveraging information from the known data-generating process, which would artificially inflate the methodology's performance metrics. Specifically, domain expertise is utilized solely for establishing temporal priors—enforcing the logical sequence where covariates and treatment assignment precede outcome observation—and for correcting directional inconsistencies in algorithms that do not inherently support temporal constraints. This restricted application notably excludes several components outlined in Section \ref{sec:identification}, including the identification of missing relationships, detection of spurious correlations, and general edge direction refinement through expert consultation. In cases where the initial causal discovery algorithms produce cyclic graphs, rather than employing the comprehensive domain knowledge-driven approach described for real-world applications, a simplified programmatic cycle-breaking procedure is implemented to maintain methodological integrity while avoiding reliance on the known underlying causal structure.

\subsection{Positivity assumption violation}
The experimental configuration for detecting and addressing positivity assumption violations employed a systematic approach based on the methodology described in Section \ref{sec:positivivity}. For violation detection, an epsilon threshold of $\epsilon = 0.001$ was established to identify regions of practical positivity violations, defined as areas where the conditional probability of treatment assignment falls below this threshold. The treatment space was partitioned into consecutive 2\% intervals to facilitate granular analysis of violation patterns across the continuous treatment domain.

To mitigate the impact of identified positivity violations, a domain-knowledge-driven data augmentation strategy was implemented. This approach leverages a fundamental assumption from the banking domain: the monotonic relationship between debt loss and repayment probability. Specifically, if a customer repays their debt under a 60\% writedown offer, it is reasonable to assume they would also repay under more favorable conditions (70\%, 80\%, ..., 100\%) where a larger portion of the debt is assumed as a loss. Therefore, a synthetic sample of this customer can be added to the dataset where the treatment values is higher than the observed and the debt is also repaid. Conversely, if a customer defaults under a 40\% writedown offer, they would likely default under less favorable conditions (30\%, 20\%, ..., 0\%) where a smaller portion of the debt is assumed as a loss. We term this monotonicity-based augmentation approach "domain-knowledge data-propagation." 

The synthetic data generation was calibrated to augment approximately 20\% of the training sample size, targeting specifically the regions identified as violating the positivity assumption. The augmentation process was applied exclusively to the training dataset to maintain the integrity of the validation split for unbiased performance assessment. Sample generation was distributed randomly across the problematic regions, ensuring balanced coverage of various violation patterns.

This strategy aligns with the framework's broader philosophy outlined in Section \ref{sec:positivivity}, which acknowledges that the selection of positivity violation remediation approaches must balance practical constraints against methodological rigor. Given the inherent limitations of conventional strategies in real-world observational data—where restricting analysis to positivity-compliant regions often results in prohibitively small samples, and limiting the adjustment set risks introducing significant confounding bias—our domain-knowledge-driven augmentation approach represents a pragmatic compromise between theoretical purity and practical utility.

\subsection{Estimation}
The experimental evaluation employed specific implementations for each estimation methodology. For the S-learner, a CatBoost classifier was utilized.

The AIPTW implementation comprised three distinct components: (i) a Generalized Propensity Score model utilizing a CatBoost regressor combined with kernel density estimation as detailed in Section \ref{sec:positivivity}, (ii) an outcome model employing a CatBoost classifier, and (iii) a final model consisting of a CatBoost regressor fitted on the pseudo-outcomes generated by the methodology, followed by logistic regression calibration to ensure well-calibrated probability estimates. To mitigate potential overfitting, all AIPTW component models were trained using out-of-sample predictions through cross-validation.

\section{Variables Definitions}

\begin{align*}
    X_{1i} &= \text{years since default} \\
    X_{2i} &= \text{default debt amount} \\
    X_{3i} &= \text{number of loans} \\
    X_{4i} &= \text{external debt} \\
    X_{5i} &= \text{number of cards} \\
    X_{6i} &= \text{loss given default} \\
    X_{7i} &= \text{number of refinances} \\
    X_{8i} &= \text{customer history length} \\
    X_{9i} &= \text{number of accounts} \\
    X_{10i} &= \text{months since first payment} \\
\end{align*}

*all features are standardized before applying the formula

\subsection{Treatment assignment formula}
The treatment value for each individual $i$ is generated through:

\begin{equation}
T_{i} = \operatorname{clip}\left(100\cdot\frac{1}{1+e^{-\theta^{\top}X_{i}}} + \epsilon_{i},\ 0,\ 100\right)
\label{eq:treatment}
\end{equation}

where the linear predictor $\theta^\top X_i$ is defined as:
\begin{equation}
    \begin{split}
        \theta^{\top}X_i = 0.5X_{1i} + 0.4\log(1+X_{2i}) + 0.3X_{3i} + 0.3\log(1+X_{4i}) + \\ 0.2X_{5i} + 0.3X_{6i} + 0.2X_{7i} + 0.1X_{1i}\log(1+X_{2i}) + 0.1X_{3i}^2
    \end{split}
\end{equation}

and $\epsilon_i$ follows a truncated normal distribution:
\begin{equation}
\epsilon_{i} \sim \mathcal{T}\mathcal{N}(0,\sigma^{2}=25,a=0,b=100)
\end{equation}

The description of each variable $X_{ji}$ can be found in the Appendix's variables definition.

\subsection{Outcome generation formula}
The outcome generation process comprises two distinct stages: an initial probability computation followed by a structured sampling procedure that incorporates domain-specific monotonicity constraints. For each individual $i$, the base probability of a positive outcome given treatment $t$ is initially modeled as:
\begin{equation}
    P(Y_i = 1|T = t, X_i) = \begin{cases}
        0 & \text{if } t = 0 \\
        1 & \text{if } t = 100 \\
        \left(\frac{t}{100}\right)^{\exp(\eta_i)}  & \text{otherwise}
    \end{cases}
    \label{eq:outcome}
\end{equation}
where the individual-specific coefficient $\eta_i$ is computed through:
\begin{equation}
    \begin{split}
        \eta_i = & 0.6X_{1i} + 0.5\log(1+X_{2i}) + 0.5X_{3i} + 0.4\log(1+X_{4i}) + 0.3X_{5i} - \\
        & 0.4X_{8i} - 0.3X_{9i} - 0.2X_{10i} + 0.1X_{1i}\log(1+X_{2i}) + 0.1X_{3i}^2
    \end{split}
\end{equation}
where the description of each variable $X_{ji}$ can be found in the Appendix's variables definition.
These base probabilities undergo transformation through a binomial sampling process. The procedure first applies probability bounding:
\begin{equation}
\tilde{p}_i = \operatorname{clip}(P(Y_i = 1|T = t, X_i), 0, 1)
\label{eq:prob_clip}
\end{equation}
followed by binomial sampling to determine the realized outcome:
\begin{equation}
Y_i \sim \text{Bernoulli}(\tilde{p}_i)
\label{eq:binom_sample}
\end{equation}
The final conditional average dose-response curves incorporate domain-specific monotonicity assumptions through the following formulation, where $t_{obs}$ represents the observed treatment level and $s$ the intervention :
\begin{equation}
P(Y_i = 1|T = s, X_i) = \begin{cases}
0 & \text{if } Y_i = 0 \text{ and } s \leq t_{obs} \\
1 & \text{if } Y_i = 1 \text{ and } s \geq t_{obs} \\
\tilde{p}_i & \text{otherwise}
\end{cases}
\label{eq:monotone_final}
\end{equation}

\section{Figures}\label{sec:app-figures}

\begin{figure}[H]
    \centering
    \includegraphics[width=\linewidth]{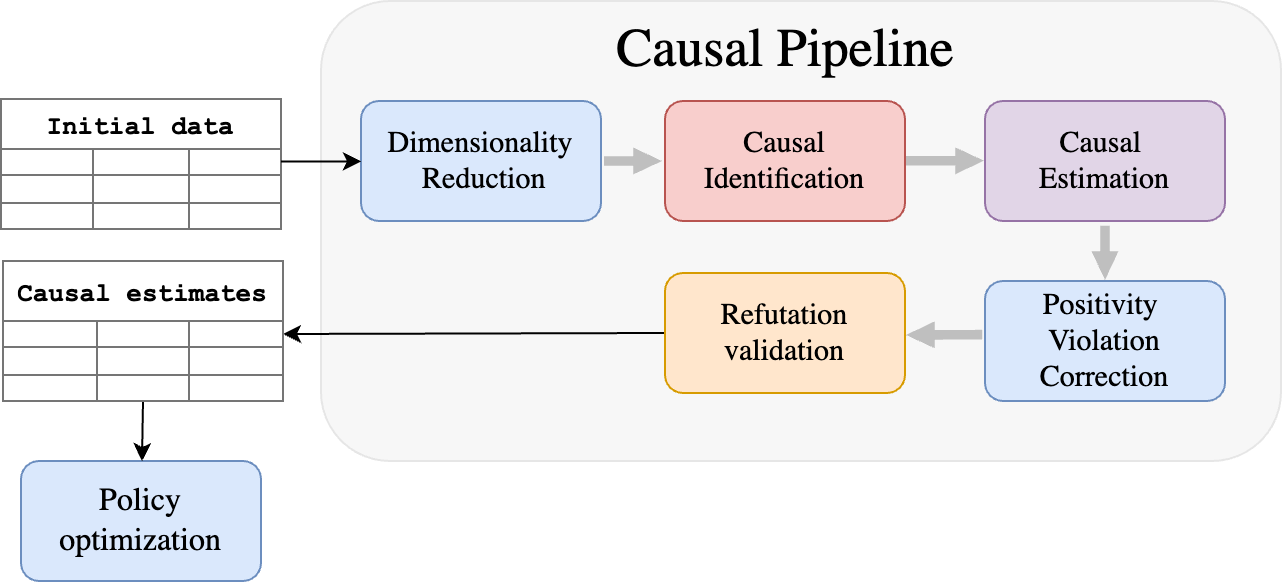}
    \caption{Diagram of the proposed causal pipeline.}
    \label{fig:pipelinediagram}
\end{figure}

\begin{figure}[H]
    \centering
    \includegraphics[width=0.4\textwidth]{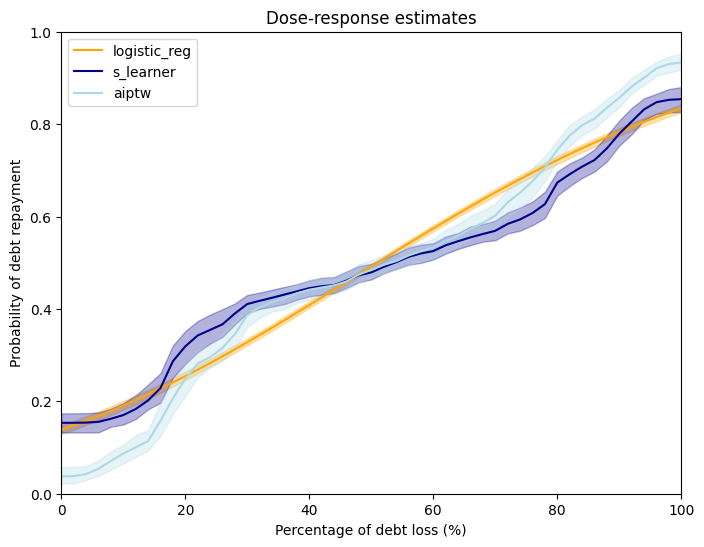}
    \caption{Estimates dose-response curves.}
    \label{fig:figure4A}
\end{figure}

\begin{figure}[H]
    \centering
    \includegraphics[width=0.7\columnwidth]{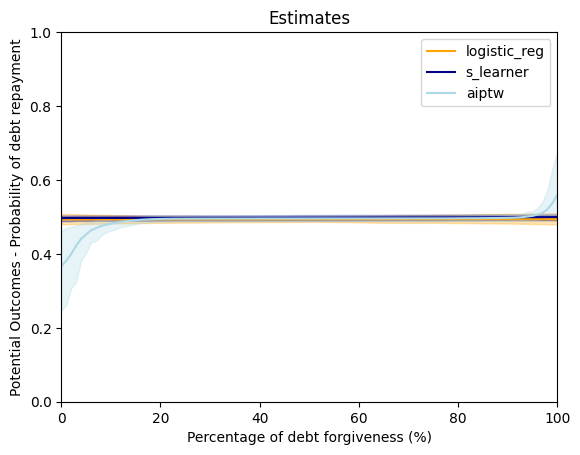}
    \caption{Placebo Treatment Replacement results}
    \label{fig:figure5A}
\end{figure}

\begin{figure}[H]
    \centering
    \includegraphics[width=0.7\columnwidth]{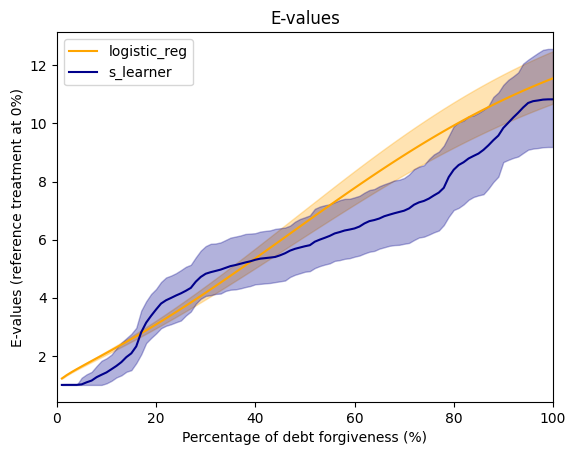}
    \caption{E-values sensitivity analysis results}
    \label{fig:figure10A}
\end{figure}

\begin{figure}[H]
    \centering
    \includegraphics[width=0.7\columnwidth]{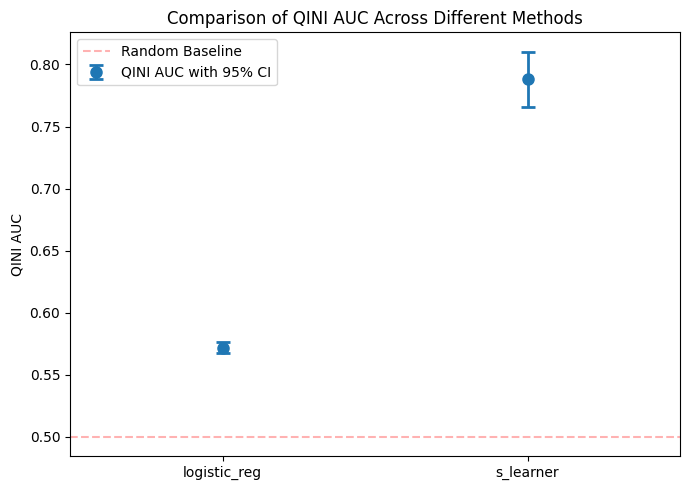}
    \caption{AUC Cumulative gain curves results}
    \label{fig:figure11A}
\end{figure}

\begin{figure}[H]
    \centering
    \includegraphics[width=0.3\textwidth]{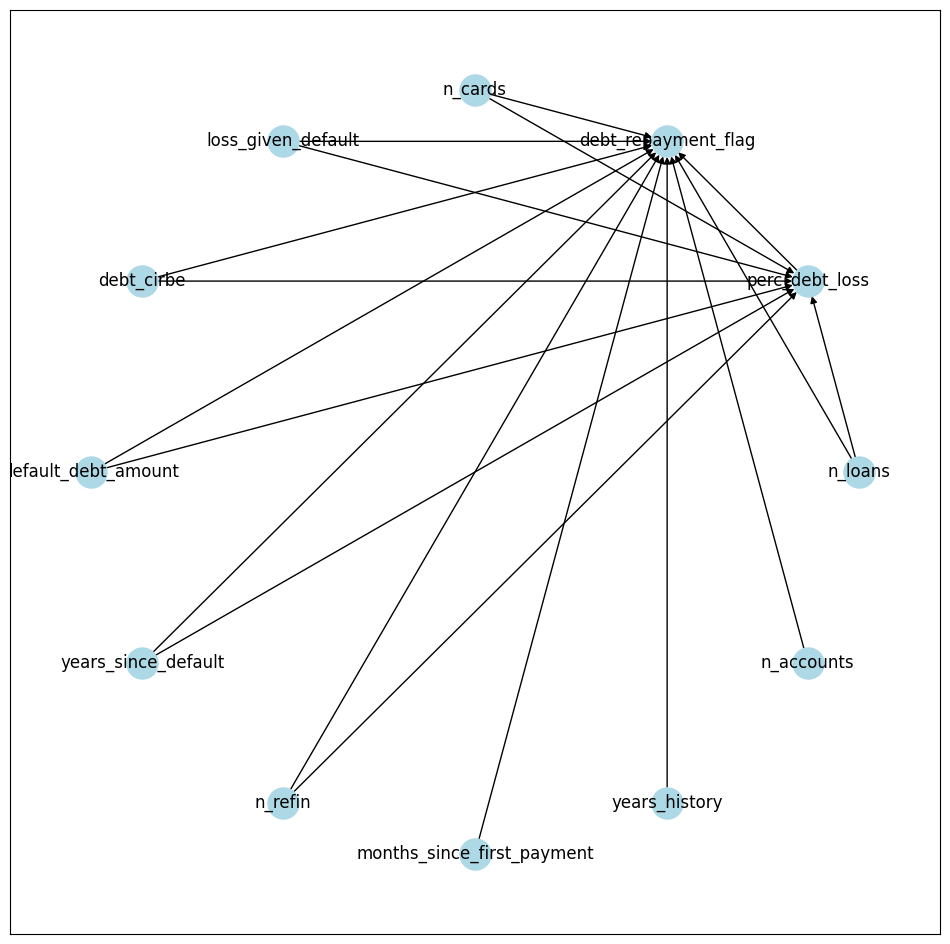}
    \caption{Estimated causal DAG.}
    \label{fig:figure7A}
\end{figure}

\vspace{-2em}

\begin{figure}[H]
    \centering
    \includegraphics[width=0.3\textwidth]{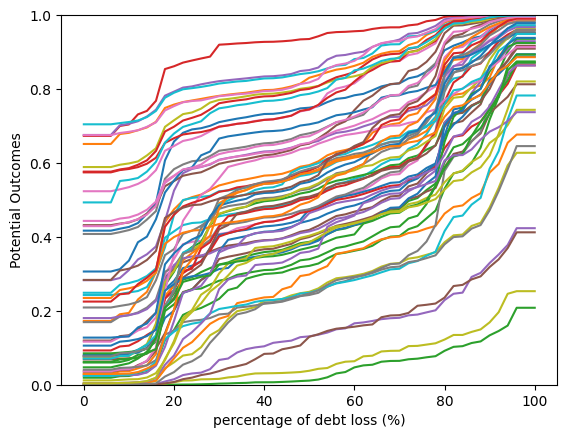}
    \caption{Individual dose-response estimated curves}
    \label{fig:figure8A}
\end{figure}

\begin{figure}[H]
    \centering
    \includegraphics[width=0.4\textwidth]{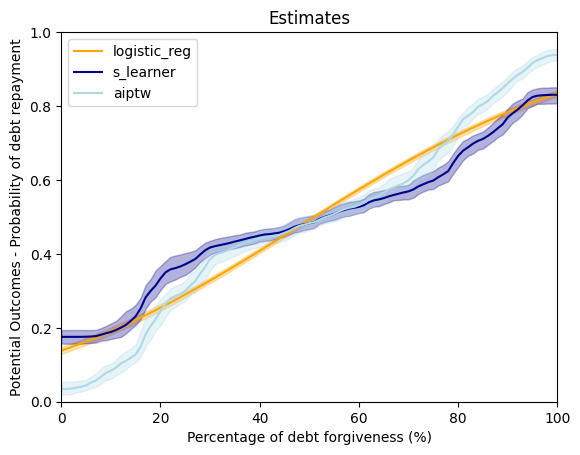}
    \caption{Random Common Cause test results.}
    \label{fig:figure6A}
\end{figure}

\begin{figure}[H]
    \centering
    \includegraphics[width=0.4\textwidth]{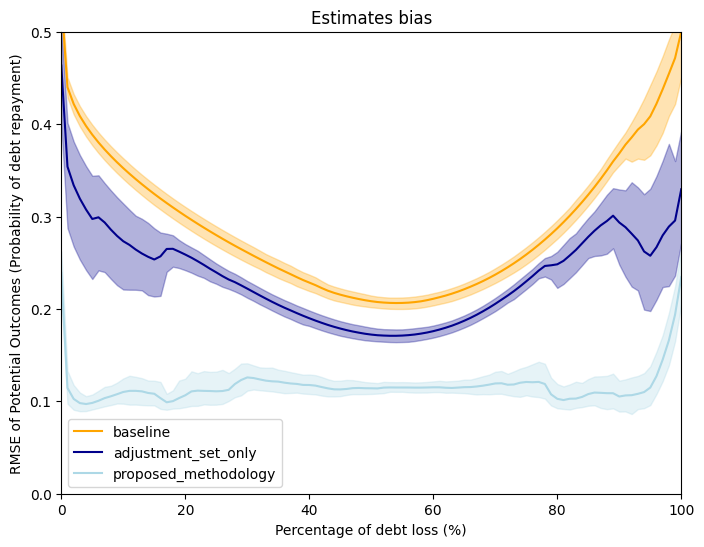}
    \caption{Estimates bias from adjustment set, baseline pipeline and proposed methodology across debt loss percentages}
    \label{fig:figure1A}
\end{figure}

\end{document}